\documentclass{PoS}

\title{The ALICE TPC: Optimization of the Performance in Run 2 and Developments for the Future}

\ShortTitle{The ALICE TPC: Performance in Run 2 and future developments}

\author{\speaker{Ernst Hellb\"{a}r} for the ALICE collaboration\\
        Goethe-Universit\"{a}t Frankfurt am Main\\
        E-mail: \email{hellbaer@ikf.uni-frankfurt.de}}

\abstract{The Time Projection Chamber is the main tracking and particle identification detector of the \mbox{ALICE} experiment. The high luminosities delivered by the CERN LHC in Run 2 (2015--2018) posed new challenges in terms of detector performance and efficiency. The hardware components and calibration software were optimized and further developed to meet those challenges and maximize the quality of the data. In addition, unexpectedly large local distortions of the drift paths of ionization electrons have been observed at the edges of specific readout chambers. These distortions are caused by ions which originate at the readout chambers, leading to local space-charge accumulation in the drift volume of the TPC. A new calibration procedure has been developed to correct for the space-charge distortions. Extensive studies have been performed to understand the origin of the space charge as well as to find a way to effectively mitigate the effect. For Run 3 starting in 2021, the new readout chambers of the upgraded TPC will be based on Gas Electron Multipliers. This implies an intrinsic backflow of ions which leads to large space-charge distortions in most of the TPC drift volume at the highest luminosities. The simulation and calibration of the space-charge effect are a major part of the new detector software framework.
}

\FullConference{7th Annual Conference on Large Hadron Collider Physics - LHCP2019\\
		20-25 May, 2019\\
		Puebla, Mexico}

\begin{document}

\section{Introduction}
The Time Projection Chamber (TPC) of the ALICE experiment is a cylindrical large-scale gaseous detector. The reconstruction of particle trajectories and the identification of these particles are its main tasks. The latter is performed by measurement of the energy loss via ionization in the detector gas and the determination of the momentum via the curvature of the trajectories in the magnetic field of 0.5 T applied by the solenoid magnet surrounding the ALICE central detectors. A detailed description of the TPC is given in \cite{ALICETPC}.

Within the drift volume of the TPC, an electric field of 400 V/cm is achieved by applying \mbox{$-100$ kV} at the central electrode. The finely segmented field cage at the inner and outer wall of the TPC provides a very high level of homogeneity of the drift field. The readout at the end caps is divided into 18 trapezoidal sectors in $\varphi$. They are referred to as sectors 0--17 on the A side and sectors 18--35 on the C side. In each sector, there is an inner (outer) readout chamber called IROC (OROC) with an active area between 84.8 cm and 132.1 cm (134.6 cm and 246.6 cm) in radial direction. The readout chambers consist of Multi-Wire Proportional Chambers (MWPCs) with pad readout and a gating grid. For Run 2, the gas mixture was changed from a Neon-based to an Argon-based mixture to guarantee the stability of the readout chambers. Furthermore, the readout control unit was replaced to increase the readout rate in Pb--Pb collisions by a factor of two.

In order to fully exploit the capabilities of the LHC in Run 3, an upgrade of the TPC readout is mandatory \cite{UpgradeTDR}. Instead of MWPCs, the new readout chambers consist of stacks of four Gas Electron Multiplier (GEM) foils, which allow continuous readout of Pb--Pb collisions at 50 kHz interaction rate (IR). As the amplification process happens inside the holes of the GEMs and there is no additional gating grid anymore, a fraction of positive ions (space charge) naturally escapes into the drift volume (ion backflow). The space charge distorts the electric field and, therefore, the measured position of the ionization electrons, leading to so-called space-charge distortions. For the final configuration of the TPC upgrade, the ion backflow is minimized to 1\% while preserving the energy resolution of the MWPCs. Therefore, at a gain factor of 2000 to be used in the Ne-CO$_{2}$-N$_{2}$ gas mixture, $\varepsilon = 20$ positive ions will enter the drift volume for each ionization electron.

\section{Space-charge distortions in Run 2}
At the high interaction rates delivered in Run 2, unexpected large space-charge distortions have been observed at very specific regions of the TPC. They appear right at the boundaries between certain IROCs and deflect the ionization electrons towards these boundaries, leading to a bias in the measured space-point position and effectively decreasing the active readout area. The size of the distortions increases linearly with the drift length of the electrons and non-trivially with the collision rate, implying that the space charge is created by gas amplification at the readout chambers.

\begin{figure}[t]
  \includegraphics[width=\textwidth]{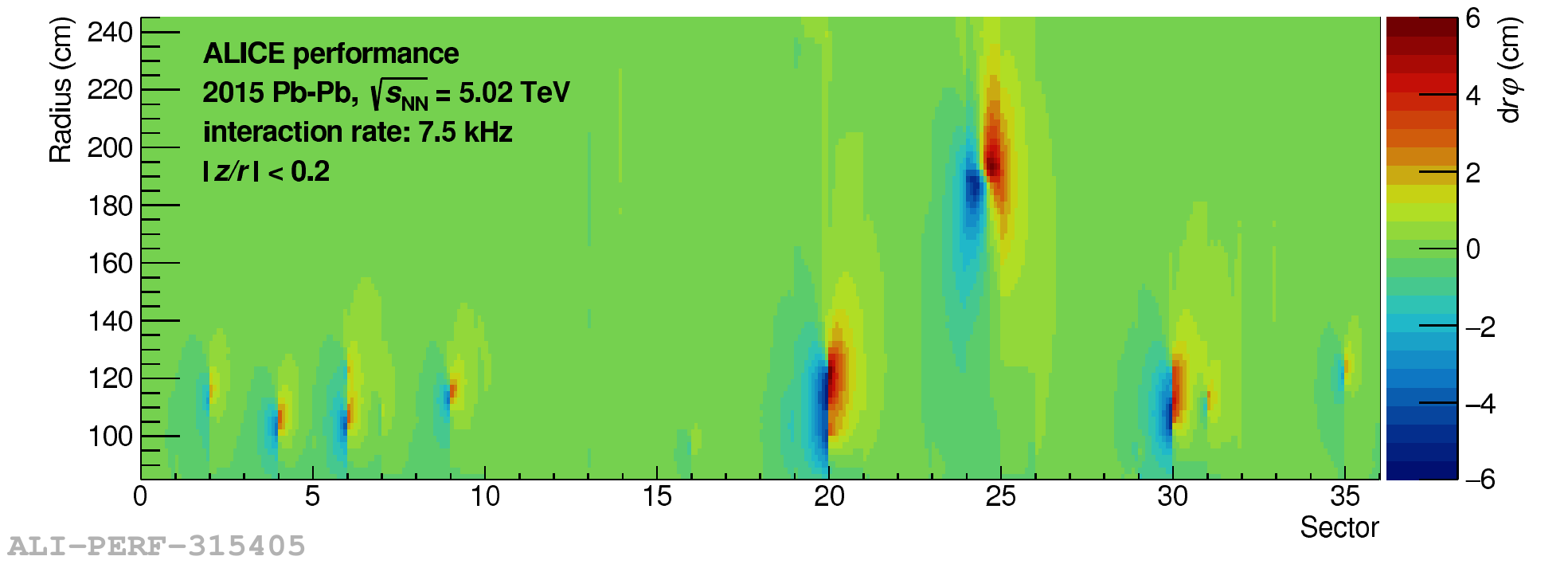}
  \caption{The measured space-charge distortions $\mathrm{d}r\varphi$ (color) as a function of the radius and the TPC sector coordinate for $|z/r| < 0.2$ in 2015 Pb--Pb data at an interaction rate of 7.5 kHz.}
  \label{fig1}
\end{figure}

Figure \ref{fig1} shows the space-charge distortions in azimuthal direction $\mathrm{d}r\varphi$ as a function of the radius and the azimuthal position in terms of the TPC sector coordinate, measured in Pb--Pb collisions at 7.5 kHz in 2015. As the space-charge distortions depend linearly on the drift distance of the electrons, the data for the largest drift distances ($|z/r|<0.2$) are selected. While there are no space-charge distortions in most of the TPC volume, at some of the IROC boundaries they reach up to 2 cm on the A side and 6 cm on the C side. The location and topology of the distortions indicate that the space charge originates at the chamber edges rather than by ion leakage through the gating grid, which will be further discussed in section \ref{sec:origin}. Moreover, the full functionality of the gating grid in all IROCs has been validated. Large space-charge distortions are also observed over the full width of one of the OROCs (sector 24) on the C side. In this case, positive ions from the amplification region enter the drift volume due to a pair of floating gating grid wires.

\subsection{Correction with external detectors}
\label{sec:correctionRUN2}
In order to compensate for the effect of the space-charge distortions, the position of each TPC cluster is corrected using a three dimensional correction map. It is obtained taking a refit of the tracklets from the Inner Tracking System (ITS), Transition Radiation Detector (TRD) and Time-Of-Flight detector (TOF) as a reference for the true track position. Dividing the full TPC volume in small voxels, the 3D distortion vectors ($\mathrm{d}r$, $\mathrm{d}r\varphi$, $\mathrm{d}z$) are calculated by comparison of the position of the distorted TPC clusters and the reference track interpolation. A smooth Chebyshev parameterization of the distortion vectors provides the final correction map.

One correction map is created for time intervals between 20 and 40 minutes, which allows a precise correction of the average space-charge distortions over the given time interval. However, the small spatial extent of the space charge leads to relative fluctuations of the space-charge distortions of about 30\% on time scales much smaller than one time interval, which makes it difficult to fully correct for the fluctuations with this procedure. An additional error is added to the covariance matrix to account for the space-charge density fluctuations.

\begin{figure}[t]
  \includegraphics[width=\textwidth]{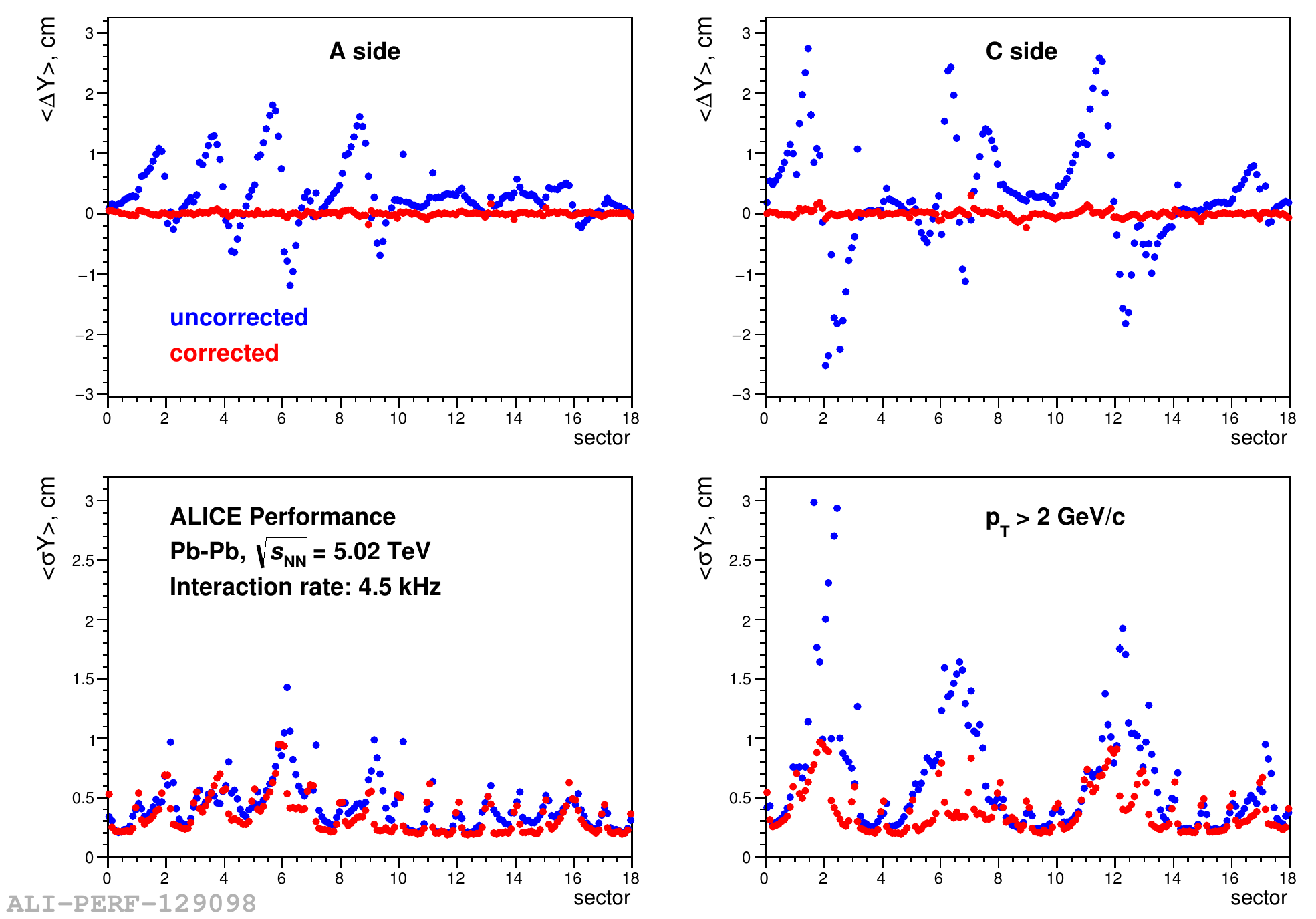}
  \caption{The track residuals to the vertex (top row) and the resolution (bottom row) are shown as a function of the TPC sector on the A (left) and C (right) side for uncorrected (blue) and corrected (red) data taken during the Pb--Pb period in 2015.}
  \label{fig2}
\end{figure}

The track residuals to the vertex and its resolution in the transverse plane are shown in figure \ref{fig2} for the uncorrected and corrected data. While the average distortions are sufficiently corrected, the resolution is clearly deteriorated in the affected regions due to the space-charge density fluctuations.

\subsection{Origin of the space charge}
\label{sec:origin}
Extensive studies have been performed to identify the exact origin of the space charge in the IROCs. The measured distortions are fitted with an analytical model. The space-charge density is assumed to have the shape of a thin line so that the resulting electric field vector can be calculated analytically with the line-charge density, its position and the distance to it in $r$ and $r\varphi$ as free parameters. Knowing the electric field, the distortions can be derived solving the Langevin equation analytically up to second order \cite{aliceint016}, allowing to use the electric field formula as a fit function. The locations of the line charges are extracted from the results of the fits for the full Pb--Pb period of 2015. The radial position differs for each sector but is constant in time. The position in $r\varphi$ is also constant in time and for each affected sector, it is inside the gap of 3 mm between two chambers.

An analysis of the cluster occupancies $N_{\mathrm{cl}}$ in high-IR data with big distortions and low-IR data without distortions gives the same conclusion. The ratio $\frac{N_{\mathrm{cl}}(\mathrm{high\textrm{-}IR})}{N_{\mathrm{cl}}(\mathrm{low\textrm{-}IR})}$ is proportional to the derivative of the distortions and provides a very precise measurement of their location due to their big gradient. The distribution of the occupancy ratio in $r\varphi$ indicates that the space charge originates inside the gap between two chambers as it is centered around that region, agreeing with the results of the analytical model.

An inspection of the MWPCs after their extraction during the upgrade of the TPC in 2019 confirmed that assumption. Single anode wire tips were sticking into, or even out of, the insulation material at the outer edge of the affected IROCs, leading to gas amplification of ionization electrons which enter the gap and the production of positive ions which then drift into the TPC volume.

\subsection{Mitigation of the space-charge distortions}
The cover electrode has been placed on top of the wire ledges and is set to a nominal potential of $-180$ V to match the potential of the drift field, thus minimizing the field distortions at the edges of the active area. As the same effect is achieved at the outer edges of the chambers, ionization electrons enter the region of the gap between two chambers where the space charge is produced by gas amplification. The amount of electrons entering the gap region can be reduced by changing the voltage at the cover electrode to more positive values, which has been studied in electrostatic simulations and in special test runs. It is found that the space-charge distortions at the IROC boundaries can be reduced by a factor of 2.5 to 10 by applying $+180$ V at the cover electrodes while limiting the static distortions at the chamber edges to a couple of millimeters. As the space-charge distortion fluctuations scale with the distortions themselves, they are also significantly decreased.

The space-charge distortions in the OROC of sector 24 due to two floating gating grid wires are also mitigated by increasing the potential difference which is applied to adjacent gating grid wires to close the gating grid. Furthermore, the high voltage of the anode wires is reduced by 50 V which decreases the gain, and therefore the amount of positive ions, by about 50\% without substantially affecting the number of clusters above threshold. In total, the space-charge distortions decrease by about a factor of four in the OROC.

\begin{figure}[t]
  \includegraphics[width=\textwidth]{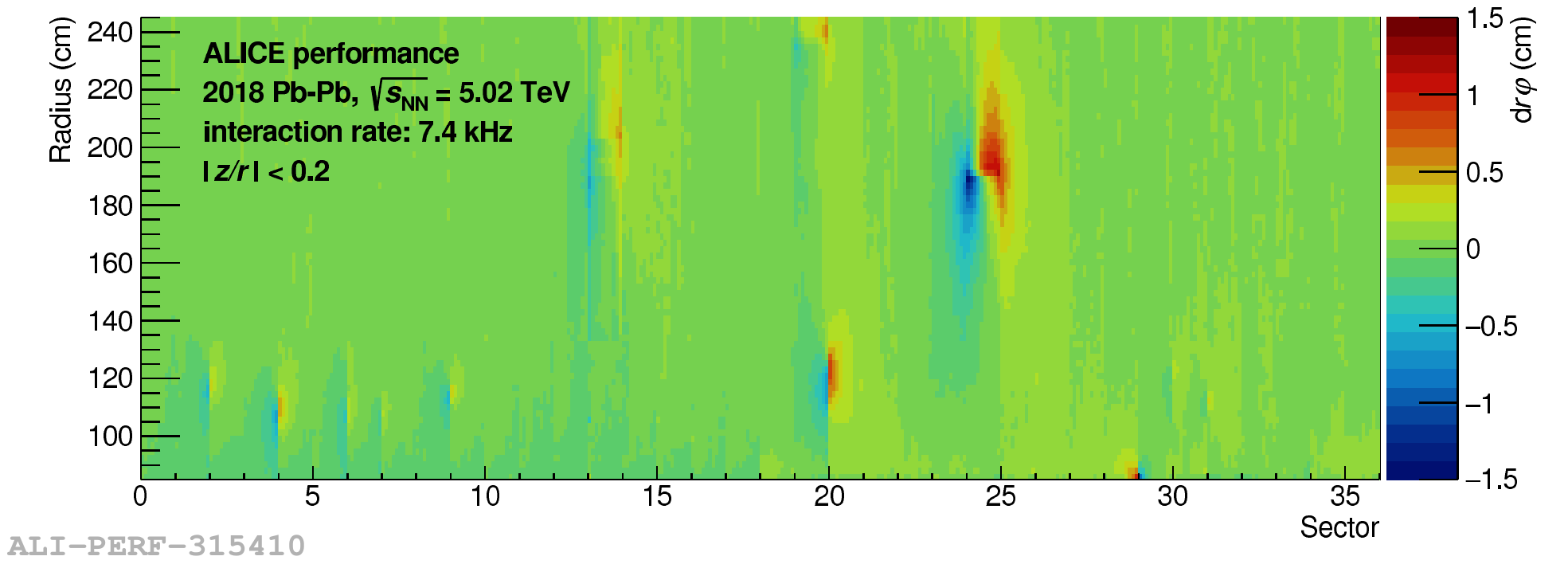}
  \caption{The space-charge distortions $\mathrm{d}r\varphi$ (color) as a function of the radius and the TPC sector for $|z/r| < 0.2$ in 2018 Pb--Pb data at an interaction rate of 7.4 kHz. Modified voltage settings are used in the IROCs and the OROC of sector 24 to mitigate the space-charge distortions.}
  \label{fig3}
\end{figure}

The final results of the mitigation of space-charge distortions are presented in figure \ref{fig3}. By applying the new voltage settings in the IROCs and the OROC of sector 24 in the last Pb--Pb period of Run 2 in 2018, the distortions are reduced to below 1 cm in the IROCs and 1.5 cm in the OROC.

\section{Space-charge distortions in Run 3}
\begin{figure}[t]
  \includegraphics[width=0.49\textwidth]{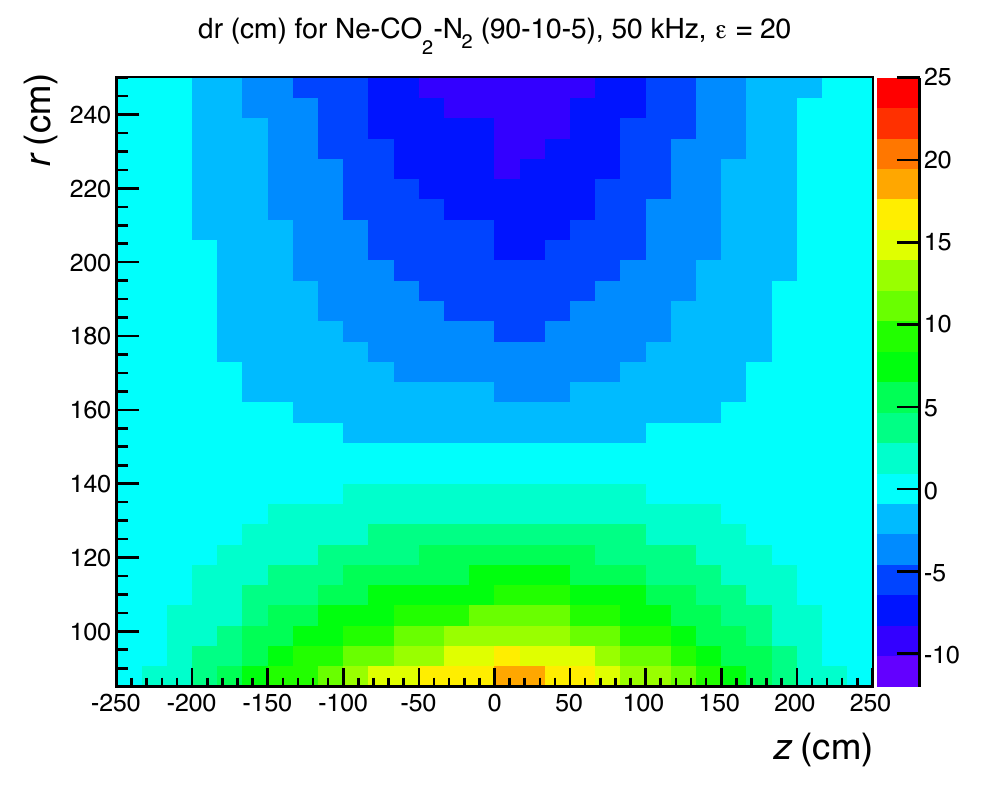}
  \includegraphics[width=0.49\textwidth]{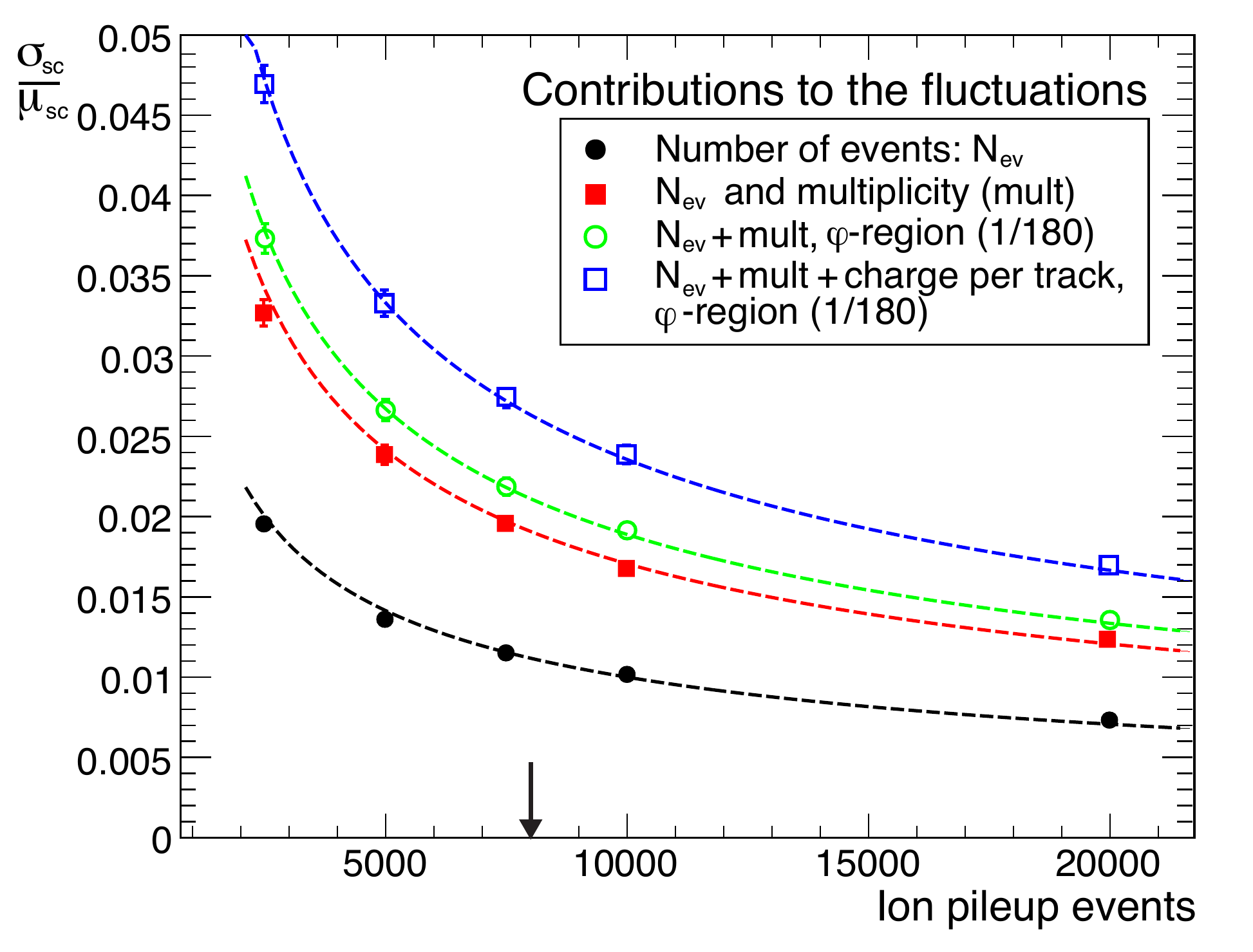}
  \caption{\textit{Left:} Radial space-charge distortions $\mathrm{d}r$ (color) as a function of the radius $r$ and the $z$ position, expected for $\varepsilon = 20$ in Pb--Pb collisions at 50 kHz. \textit{Right:} Different contributions to the space-charge density fluctuations at 50 kHz Pb--Pb collisions from equation \ref{eq:fluc} (dashed lines) and toy Monte-Carlo simulations (markers).}
  \label{fig4}
\end{figure}

Although extensive R\&D has been carried out to minimize the ion backflow as much as possible for the upgraded TPC, there will still be a significant amount of space charge in the drift volume. At ion drift times of 160--200 ms and an interaction rate of 50 kHz, ions from 8000--10000 pile-up events will contribute on average to the space-charge density at any given moment. The resulting space-charge distortions in radial direction expected at 50 kHz of Pb--Pb collisions are shown in the left panel of figure \ref{fig4}. The electrons are deflected towards the radial center of the drift volume and the absolute value of the distortions increases towards the inner and outer wall, reaching a maximum of 20 cm at the inner wall close to the central electrode. Considering the size of the expected distortions, the space-charge density fluctuations become significant with respect to the intrinsic track resolution of $\mathcal{O}(200 \ \mu\mathrm{m})$. The relative space-charge density fluctuations can be calculated analytically by
\begin{equation}
  \label{eq:fluc}
  \frac{\sigma_{\mathrm{SC}}}{\mu_{\mathrm{SC}}} = \frac{1}{\sqrt{N^{\mathrm{ion}}_{\mathrm{pileup}}}} \sqrt{1 + \left( \frac{\sigma_{N_{\mathrm{mult}}}}{\mu_{N_{\mathrm{mult}}}} \right) ^2 + \frac{1}{F\mu_{N_{\mathrm{mult}}}} \left( 1 + \left( \frac{\sigma_{Q_{\mathrm{track}}}}{\mu_{Q_{\mathrm{track}}}} \right) ^2 \right) } \; ,
\end{equation}
where $N^{\mathrm{ion}}_{\mathrm{pileup}}$ are the number of pile-up events, $\frac{\sigma_{N_{\mathrm{mult}}}}{\mu_{N_{\mathrm{mult}}}}$ is the relative RMS of the track multiplicity distribution, $\frac{\sigma_{Q_{\mathrm{track}}}}{\mu_{Q_{\mathrm{track}}}}$ is the relative variation of the ionization of a single track and $F$ is a geometrical factor describing the range over which the space-charge density fluctuations are relevant for the distortions. The right panel in figure \ref{fig4} shows the relative space-charge density fluctuations according to equation \ref{eq:fluc} together with the results of toy Monte-Carlo simulations for Pb--Pb collisions. For the expected number of pile-up events at 50 kHz, the total fluctuations are of the order of 3\%.

\subsection{Calibration strategy}
A sophisticated calibration of the data is foreseen for Run 3 in order to fully correct for the space-charge distortions and fluctuations without major loss in cluster and track resolution. The calibration will be performed in two stages. A first correction is applied online before the tracking. The TPC clusters are corrected by a stored correction map which is obtained from previous data using the ITS, TRD and TOF interpolation method (section \ref{sec:correctionRUN2}). It is updated to the current average charged particle density in time intervals of $\mathcal{O}\mathrm{(min)}$ by scaling with the signals at the readout chambers integrated over the ion drift time (integrated digital currents). With the average distortions and part of the fluctuations corrected, the online tracking can be performed with residual distortions of $\mathcal{O}\mathrm{(mm)}$. During the second stage, the space-charge density fluctuations in space and time are corrected and the intrinsic track resolution of the TPC is restored. A high-resolution correction map is calculated from the data using ITS, TRD and TOF reference tracks. In order to fully correct the fluctuations, the correction has to be applied for time intervals of about 5 ms, which is done by scaling the high-resolution map with the digital currents.

\section{Summary and Outlook}
Several components of the ALICE TPC were replaced and optimized to further improve its performance in Run 2, e.g. the gas mixture and the readout control unit. Unexpected local space-charge distortions have been observed at specific sector boundaries of the IROCs. A new calibration procedure relying on the external detectors ITS, TRD and TOF has been developed to correct the data. The origin of the space charge has been identified. It is created in the gap between two chambers by gas amplification at anode wire tips sticking out into the gap. For the last Pb--Pb period of Run 2, the space-charge distortions and distortion fluctuations could be mitigated by adjusting the voltage at the cover electrodes to $+180$ V, recovering the intrinsic track resolution of the TPC after calibration.

For the upgraded TPC, even bigger space-charge distortions are expected in a large part of the drift volume. A challenging calibration procedure based on the developments in Run 2 is foreseen to restore the intrinsic track resolution. A number of studies have been performed and developments have been started to determine the requirements for the calibration and to validate it. These include an implementation of a realistic propagation of space charge through the drift volume and the application of convolutional neural networks for the calculation of the distortions.

\end{document}